\documentclass[aps,prL,twocolumn,superscriptaddress]{revtex4-1}
\usepackage{graphicx}
\usepackage{dcolumn}
\usepackage{bm}
\usepackage{amsmath}
\usepackage[english]{babel}
\usepackage[utf8]{inputenc}
\usepackage{verbatim}
\usepackage{nicefrac}
\newcommand{\RomanNumeralCaps}[1]
    {\MakeUppercase{\romannumeral #1}}

\def\nobreakbefore{%
  \relax\ifvmode\else
    \ifhmode
      \ifdim\lastskip > 0pt\relax
        \unskip\nobreakspace
      \fi
    \fi
  \fi
}
\let\oldcite\cite
\renewcommand\cite{\nobreakbefore\oldcite}

\usepackage{xcolor}
\usepackage{upgreek}

\usepackage{layouts}
\usepackage{braket,mleftright}
\mleftright
\usepackage[T1]{fontenc}

\begin{document}
	\title{Phase flip code with semiconductor spin qubits}

\author{F. van Riggelen}
	\affiliation{QuTech and Kavli Institute of Nanoscience, Delft University of Technology, Lorentzweg 1, 2628 CJ Delft, The Netherlands}
	\author{W. I. L. Lawrie}
    \affiliation{QuTech and Kavli Institute of Nanoscience, Delft University of Technology, Lorentzweg 1, 2628 CJ Delft, The Netherlands}
	\author{M.~Russ}
	\affiliation{QuTech and Kavli Institute of Nanoscience, Delft University of Technology, Lorentzweg 1, 2628 CJ Delft, The Netherlands}
	\author{N. W. Hendrickx}
	\affiliation{QuTech and Kavli Institute of Nanoscience, Delft University of Technology, Lorentzweg 1, 2628 CJ Delft, The Netherlands}
	\author{A. Sammak}
	\affiliation{QuTech and Netherlands Organization for Applied Scientific Research (TNO), Stieltjesweg 1 2628 CK Delft, The Netherlands}
	\author{M. Rispler}
	\affiliation{QuTech, Delft University of Technology, Lorentzweg 1, 2628 CJ Delft, The Netherlands}
	\affiliation{JARA Institute for Quantum Information, Forschungszentrum Jülich GmbH, 52428 Jülich, Germany}
	\author{B. M. Terhal}
	\affiliation{QuTech, Delft University of Technology, Lorentzweg 1, 2628 CJ Delft, The Netherlands}
	\affiliation{JARA Institute for Quantum Information, Forschungszentrum Jülich GmbH, 52428 Jülich, Germany}
	\affiliation{EEMCS, Delft University of Technology, Mekelweg 4, 2628 CD Delft, The Netherlands}
	\author{G. Scappucci}
	\affiliation{QuTech and Kavli Institute of Nanoscience, Delft University of Technology, Lorentzweg 1, 2628 CJ Delft, The Netherlands}
	\author{M. Veldhorst}
	\email{m.veldhorst@tudelft.nl}
	\affiliation{QuTech and Kavli Institute of Nanoscience, Delft University of Technology, Lorentzweg 1, 2628 CJ Delft, The Netherlands}	
	
	\begin{abstract}
    The fault-tolerant operation of logical qubits is an important requirement for realizing a universal quantum computer. Spin qubits based on quantum dots have great potential to be scaled to large numbers because of their compatibility with standard semiconductor manufacturing. Here, we show that a quantum error correction code can be implemented using a four-qubit array in germanium. We demonstrate a resonant SWAP gate and by combining controlled-Z and controlled-$\text{S}^{-1}$ gates we construct a Toffoli-like three-qubit gate. We execute a two-qubit phase flip code and find that we can preserve the state of the data qubit by applying a refocusing pulse to the ancilla qubit. In addition, we implement a phase flip code on three qubits, making use of a Toffoli-like gate for the final correction step. Both the quality and quantity of the qubits will require significant improvement to achieve fault-tolerance. However, the capability to implement quantum error correction codes enables co-design development of quantum hardware and software, where codes tailored to the properties of spin qubits and advances in fabrication and operation can now come together to scale semiconductor quantum technology toward universal quantum computers.

	\end{abstract}
	\maketitle
	
    A universal quantum computer may be able to address a range of challenges \cite{reiher_elucidating_2017, shor_polynomial-time_1999, grover_fast_1996}, but will require many logical qubits for fault-tolerant operation \cite{preskill_reliable_1998}.
    While errors on individual qubits are unavoidable, logical qubits can be encoded in multiple physical qubits, facilitating error correction codes that preserve the quantum state \cite{terhal_quantum_2015}. There are several ways to encode logical qubits that allow for the correction of different errors using a variety of error-correction strategies. The simplest error correction codes, the bit flip code and the phase flip code \cite{cory_experimental_1998, moussa_demonstration_2011, schindler_experimental_2011, reed_realization_2012, taminiau_universal_2014}, correct for the fundamental type of error after which they are named. One of the most promising error correction codes is the surface code \cite{fowler_surface_2012}, which can correct any error affecting a sufficiently low number of qubits. Using an error correction code, however, can only help to achieve low logical error rates when all error rates on the physical qubits (of initialization, control and readout) are below a threshold, dependent on the protocol. 
    
    \begin{figure*}[!t]
	\includegraphics[width = \textwidth]{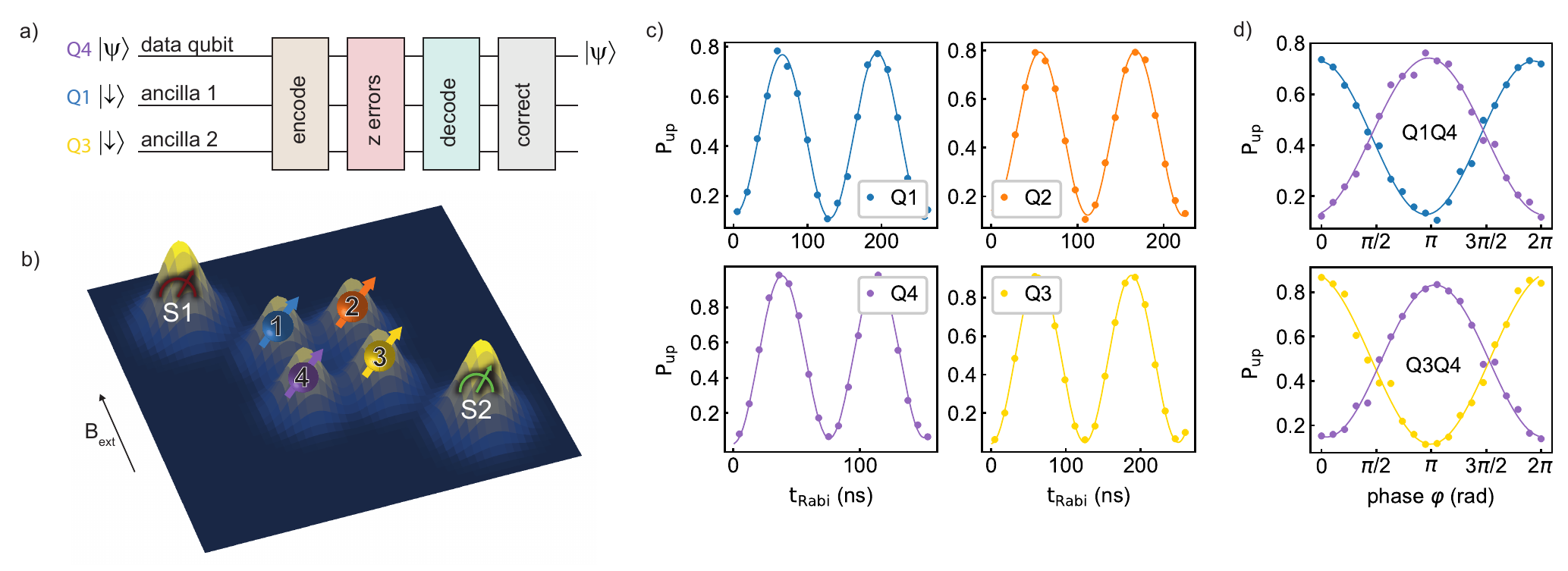}%
	\caption{\textbf{One- and two-qubit gates on a four spin qubit register.} (a) Quantum error correction circuit using one data qubit and two ancilla qubits. Qubits are encoded to a logical state (beige) in which they are resilient against single-qubit phase errors (soft red), since these errors result in distinct states of the ancilla qubits which, after the decoding (turquoise), can be used to correct the data qubit (grey). 
	(b) Schematic drawing giving an impression of the electrostatic potential of the quantum device. Using electrostatic potentials on metallic gates, four quantum dots are defined, each containing a single hole spin qubit. The qubits are indicated with a color: qubit 1 (Q1) in blue, qubit 2 (Q2) in orange, qubit 3 (Q3) in yellow and qubit 4 (Q4) in purple. The spin states are read out by spin-to-charge-conversion using latched Pauli blockade using the two charge sensors, S1 and S2, indicated in red and green respectively. (c) Spin-up probability ($\text{P}_\text{up}$) of the four qubits as a function of the duration of the applied driving pulse. Coherent Rabi-oscillations can be observed for all four qubits.
	(d) Implementation of a CPhase gate, using a Tukey shaped exchange pulse on the barrier gate. 
	This experiment is performed with (blue and yellow for target qubit Q1 or Q3 respectively) and without (purple) an $\text{X}^2$ pulse on the control qubit. The difference in acquired phase is calibrated to be $\pi$, thus implementing a CZ gate \cite{hendrickx_four-qubit_2021}.}
	\label{fig:figure_1}
    \end{figure*}
    
    The relevance of quantum error correction has spurred significant research in a multitude of platforms and exciting progress has been made in superconducting qubits \cite{rosenblum_fault-tolerant_2018, campagne-ibarcq_quantum_2020,andersen_repeated_2020, marques_logical-qubit_2021}, solid-state qubits using NV centers in diamond \cite{waldherr_quantum_2014,cramer_repeated_2016, abobeih_fault-tolerant_2021}, and trapped-ion qubits \cite{nigg_quantum_2014, egan_fault-tolerant_2021}. 
    Semiconductor qubits based on spins in quantum dots have not yet advanced to match the larger qubit counts of competing technologies \cite{rispler_towards_2020}, but important progress has been made in achieving high-fidelity operations. Fast and high-fidelity readout  \cite{harvey-collard_high-fidelity_2018,zheng_rapid_2019}, single-qubit control \cite{yoneda_quantum-dot_2018,yang_silicon_2019, hendrickx_four-qubit_2021, lawrie_simultaneous_2021}, two-qubit logic \cite{xue_quantum_2022, noiri_fast_2022, madzik_precision_2022}, and resonant three-qubit and four-qubit gates \cite{hendrickx_four-qubit_2021} have been demonstrated in separate experiments.
    
    Quantum wells in planar germanium heterostructures (Ge/SiGe) can bring together the advantages of several semiconductor quantum dots platforms \cite{scappucci_germanium_2021}. Like silicon, natural germanium contains nuclear-spin-free isotopes and can be isotopically purified \cite{itoh_high_1993, itoh_isotope_2014}. Hole states in Ge/SiGe have a low effective mass \cite{lodari_light_2019}, relaxing the fabrication requirements of nanostructures. Moreover, the strong spin-orbit interaction allows for fast and all-electric qubit operation \cite{bulaev_electric_2007, hendrickx_fast_2020, maurand_cmos_2016, watzinger_germanium_2018}. 
    Hole spin qubits in Ge/SiGe do not suffer from valley degeneracy \cite{lodari_light_2019}, which still presents a major challenge for electrons in silicon \cite{zwanenburg_silicon_2013, wuetz_atomic_2021}. Furthermore, advancements in heterostructure growth have yielded low disorder and charge noise \cite{lodari_low_2021}. These characteristics have facilitated the development of planar germanium quantum dots \cite{hendrickx_gate-controlled_2018} and quantum dot arrays  \cite{lawrie_quantum_2020}, long spin relaxation times \cite{lawrie_spin_2020}, single-hole qubits \cite{hendrickx_single-hole_2020}, singlet-triplet qubits  \cite{jirovec_singlet-triplet_2021}, two-qubit logic \cite{hendrickx_fast_2020}, and universal operation of a four-qubit germanium quantum processor  \cite{hendrickx_four-qubit_2021}. The spin-orbit coupling in germanium avoids the need to implement components such as striplines and nanomagnets, promising scalability in two dimensions \cite{van_riggelen_two-dimensional_2021, hendrickx_four-qubit_2021}, crucial for the implementation of error correction codes \cite{terhal_quantum_2015}. 
    

    Here, we perform quantum error correction on a two-by-two array of spin qubits in germanium. Like in most spin qubit platforms, hole spin qubits have long relaxation times \cite{lawrie_spin_2020}, such that the dominant type of decoherence is dephasing. We therefore focus on the implementation of a rudimentary phase flip code. In order to realize this, we implement a controlled-Z (CZ) gate, a controlled-$\text{S}^{-1}$ ($\text{CS}^{-1}$) gate and a native resonance SWAP gate \cite{sigillito_coherent_2019}. Using the CZ and $\text{CS}^{-1}$ gates we construct a Toffoli-like gate. Additionally, we show that we can coherently transfer phase information between our data and ancilla qubits and effectively implement the majority vote for error correction of the phase flip code on three qubits. 
    
    The error correction code considered here is a three-qubit phase flip code \cite{nielsen_quantum_2000}, the steps  of which are depicted in Figure \ref{fig:figure_1}a. At the start of the experiment, the data qubit could in principle hold any quantum state $\ket{\Psi }$ = $\alpha\ket{0} + \beta\ket{1}$ and both ancilla qubits start in the basis state $\ket{0}$. In the encoding step, the state of the data qubit is mapped to the ancilla qubits and the system is brought into the state $\alpha\ket{+++} + \beta\ket{---}$. After the encoding, we intentionally induce errors either by deterministically implementing a rotation around the Z axis of the Bloch sphere with angle $\varphi$ (Z($\varphi$)), a rotation Z($\varphi = \pi$) with a certain probability $p$, or by leaving the qubits idle for some time. In the decoding step, we disentangle the logical qubit, where all single phase errors lead to a unique error syndrome. In the final step of the code, a phase error is corrected. The ancillas are not measured, but the data qubit is corrected using a three-qubit gate depending on the error syndrome of the ancilla qubits \cite{ercan_measurement-free_2018}. This correction protocol is capable of correcting any phase error Z($\varphi$) on a single qubit, but it cannot correct phase errors that occur on different physical qubits simultaneously, nor can it handle errors in the encoding, decoding and correction steps.
    
    \begin{figure*}[!t]
	\includegraphics[width = \textwidth]{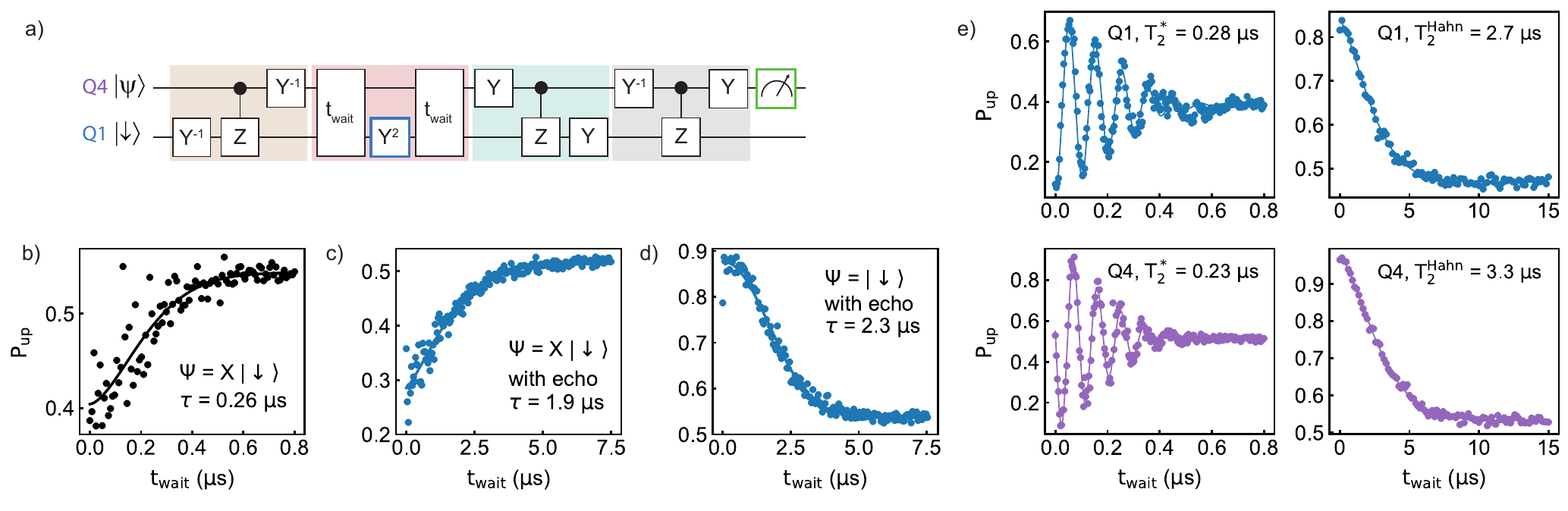}%
	\caption{\textbf{Two-qubit phase flip code.} (a) Circuit diagram. 
	The encoding, decoding and correction are implemented using a combination of Y, $\text{Y}^{-1}$ and CZ gates. By adding a wait time ($\text{t}_\text{wait}$) after the encoding, phase errors will occur due to the dephasing of the qubits. Q4 is the data qubit and Q1 the ancilla qubit. (b-d) $\text{P}_\text{up}$ as a function of $\text{t}_\text{wait}$ when executing the two-qubit phase flip code, which gives a decay time $\tau$. Results for the data qubit prepared in $\ket{\Psi }$ = X$\ket{\downarrow}$ and through single qubit gates projected to a basis state for readout, without an echo pulse $\text{Y}^2$ (b), with an echo pulse $\text{Y}^2$ (c), and with an echo pulse with the data qubit prepared to $\ket{\Psi }$ = $\ket{\downarrow}$ (d). (e) Individual qubit dephasing $\text{T}_2^*$ and coherence $\text{T}_{2}^\text{Hahn}$ times for Q1 (blue) and Q4 (purple). }
	\label{fig:figure_2}
    \end{figure*}

    The implementation of the phase flip code strongly depends on the design and properties of the quantum device. The quantum dots are defined in a strained germanium quantum well, using two layers of metallic gates and low resistance Ohmic contacts are made by diffusing aluminium contacts directly into the quantum well \cite{lawrie_quantum_2020, hendrickx_four-qubit_2021}. Figure \ref{fig:figure_1}b gives an impression of the potential landscape which is formed by applying negative voltages on four plunger gates, forming quantum dots underneath. Each quantum dot is occupied by a single hole spin. The coupling between the quantum dots is controlled by dedicated barrier gates. We construct virtual barrier and plunger gates at the software level, to independently control the detuning, on-site energy, and exchange \cite{hendrickx_four-qubit_2021}. Two additional quantum dots (S1 and S2) act as charge sensors, operated using radio frequency reflectometry for rapid readout \cite{hendrickx_four-qubit_2021}. Spin state readout is achieved using spin-to-charge-conversion in the form of latched Pauli spin blockade (PSB) \cite{studenikin_enhanced_2012, hendrickx_four-qubit_2021} (see Supplementary Information section \RomanNumeralCaps{1}). We can read out the spin state of Q1 and Q2 using S1 (readout system Q1Q2, red) and the spin state of Q3 and Q4 using S2 (readout system Q3Q4, green).
    
    An external magnetic field of 0.65 T is applied in plane of the quantum well, resulting in energy splittings of 1.393 GHz, 2.192 GHz, 2.101 GHz and 2.412 GHz for Q1, Q2, Q3 and Q4 respectively, between the spin down $\ket{\downarrow}$ (which we define to be $\ket{0}$) and spin up ($\ket{1}$). Here, we use the convention of notating an X (Y) gate as a $\pi/2$ rotation, $\text{X}^2$ ($\text{Y}^2$) as a $\pi$ rotation and $\text{X}^{-1}$ ($\text{Y}^{-1}$) as a $-\pi/2$ rotation around the $\hat{x}$ ($\hat{y}$) axis of the Bloch sphere \cite{watson_programmable_2018}. Single qubit rotations are implemented by electric dipole spin resonance and calibrated by fitting the resulting Rabi oscillations, as shown in Figure \ref{fig:figure_1}c. 
    
    The choice of two-qubit gate is also dictated by the properties of the device. Fast controlled-Z (CZ) gates \cite{veldhorst_two-qubit_2015} are possible by controlling the exchange interaction using the barrier gates \cite{hendrickx_four-qubit_2021}. Figure 1d shows the calibration of a CZ gate via a Ramsey type experiment, where we use a Tukey shaped pulse to turn the exchange on and off. Details of this experiment can be found in Supplementary Information section \RomanNumeralCaps{2}. We use CZ gates between Q1 and Q4 and between Q3 and Q4 for the entangling and disentangling in the phase flip code.

    \begin{figure*}[!t]
	\includegraphics[width = \textwidth]{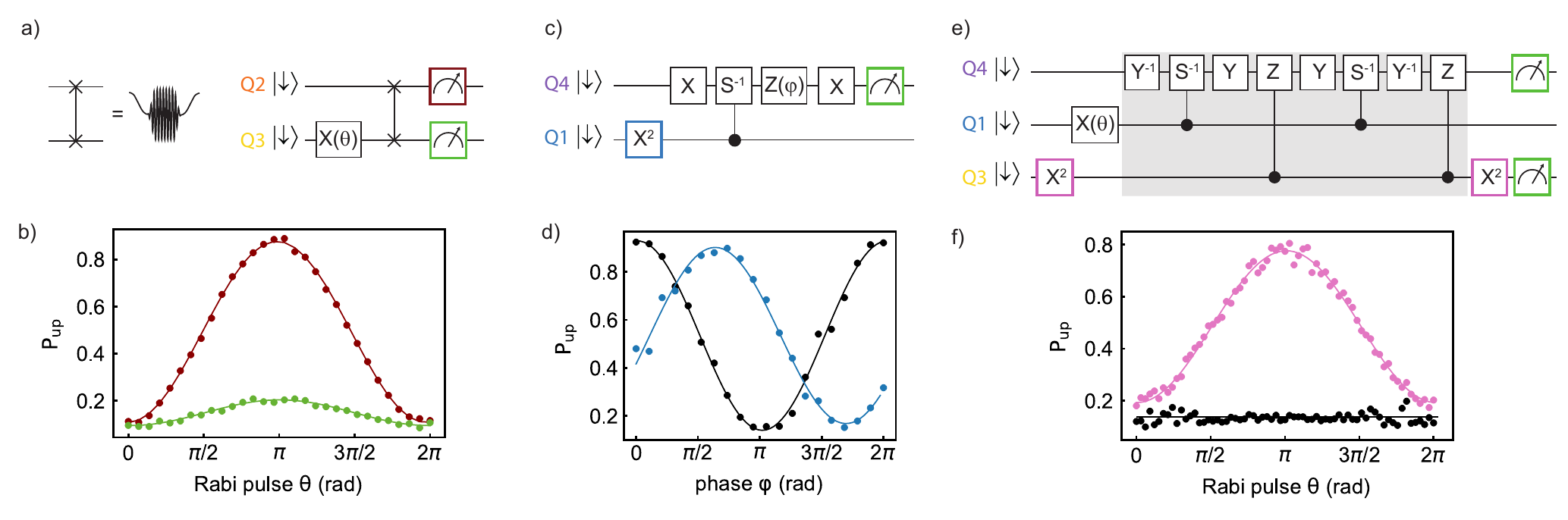}%
	\caption{\textbf{SWAP, $\text{CS}^{-1}$ and Toffoli-like gate} (a) A resonant SWAP gate is implemented by applying a Tukey shaped pulse with an oscillation superimposed to the barrier gate between Q2 and Q3. It is tested using the depicted circuit diagram. (b) To demonstrate the SWAP gate, a Rabi pulse X$(\theta)$ is applied to Q3, followed by a SWAP gate between Q2 and Q3. Then, either Q2 is read out using readout system Q1Q2 (red) or Q3 is read out using readout system Q3Q4 (green). (c) Circuit diagram of the experiment demonstrating the $\text{CS}^{-1}$ gate. (d) The $\text{CS}^{-1}$ gate is obtained by calibrating the phase difference to be $-\pi/2$, for the experiments with (blue) and without (black) a $\text{X}^2$ gate on the control qubit Q1. (e) Circuit diagram of the Toffoli-like gate (gray) composed of $\text{CS}^{-1}$ and CZ gates. (f) Demonstration of the Toffoli-like gate with target qubit Q4 and control qubits Q1 and Q3. An X$(\theta)$ pulse is applied to Q1 and the final state of Q4 is measured using readout system Q3Q4. Results of the experiment in (e) with (pink) and without (black) a preparation pulse $\text{X}^2$ on ancilla qubit Q3. 
	}
	\label{fig:figure_3}
    \end{figure*}
    
    As a stepping stone towards the three-qubit phase flip code, we first implement a two-qubit phase flip code. The two-qubit code consists of the same steps (encode, phase errors, decode, and correct) but differs from the three-qubit code in that a phase error can only be corrected on the data qubit.
    
    The compiled gate set of the two-qubit phase flip code is depicted in Figure \ref{fig:figure_2}a. We use Q4 as data qubit and Q1 as ancilla qubit. The encoding (beige) is performed by a Hadamard-CZ-Hadamard sequence \cite{nielsen_quantum_2000}, where the Hadamards are replaced by $\text{Y}^{-1}$ gates.
    The phase errors are induced by leaving the qubits idle for some time (soft red). Since this code should correct for a phase error on the data qubit, one would expect that the dephasing time of the ancilla qubit Q1 is the limiting factor. Ramsey experiments (Figure \ref{fig:figure_2}e) yield pure dephasing times ($\text{T}_2^*$) of 0.28 $\pm$ 0.1 $\upmu$s and 0.23 $\pm$ 0.1 $\upmu$s for Q1 and Q4 respectively. These are comparable to the decay time ($\tau$) of $0.26 \pm 0.01 \upmu$s corresponding to the two-qubit phase flip code, shown in Figure \ref{fig:figure_2}b. The fact that the phase errors on the ancilla qubit are limiting can be seen even more clearly when a refocusing pulse is applied to the ancilla qubit (blue box in Figure \ref{fig:figure_2}a). The result of this experiment is shown in Figure  \ref{fig:figure_2}c and gives $\tau = 1.86$ $\pm$ 0.05 $\upmu$s. We have also run this experiment with the data qubit starting in the basis state $\ket{\downarrow}$ (see Supplementary Information section \RomanNumeralCaps{3}). The result is shown in Figure \ref{fig:figure_2}d and gives $\tau = 2.31$ $\pm$ 0.02 $\upmu s$. For comparison, the results of a Hahn echo experiment are shown for both Q1 and Q4 in Figure \ref{fig:figure_2}e. We extract $\text{T}_{2}^\text{Hahn}$ = 2.72 $\pm$ 0.05 $\upmu$s and 3.26 $\pm$ 0.04 $\upmu$s for Q1 and Q4 respectively. The two-qubit phase flip code is also performed with Q3 as ancilla qubit instead of Q1 (Supplementary Information section \RomanNumeralCaps{3}). All off these experiments show that the state of the data qubit can be preserved for a longer time when a echo pulse is applied to the ancilla qubit, giving a typical decay time in the order of the $\text{T}_{2}^\text{Hahn}$ of the ancilla qubit.  

    Since we use PSB readout, we can only read out the state of an individual qubit when the state of the other qubit in the readout system is known. Therefore, when using Q4 as data qubit and Q3 as one of the ancilla qubits, it is necessary to reinitialize Q3. We enable this by performing a SWAP gate on Q3 and Q2, with Q2 initialized to the state $\ket{\downarrow}$. Implementing a diabatic SWAP gate is difficult due to the relatively large Zeeman energy difference between Q2 and Q3 \cite{meunier_efficient_2011, russ_high-fidelity_2018}. While a SWAP can be compiled from a series of CZ gates and single-qubit operations, here we implement a resonant SWAP \cite{sigillito_coherent_2019} by applying an electric pulse as depicted in Figure \ref{fig:figure_3}a to the barrier gate. This pulse is an oscillating exchange pulse, resonant with the difference in Zeeman energies of Q2 and Q3, superimposed on a Tukey shaped pulse (see Supplementary Information section \RomanNumeralCaps{4} for details on calibration). Figure \ref{fig:figure_3}a shows the circuit diagram to demonstrate resetting the state of Q3 using the SWAP. The result is shown in Figure \ref{fig:figure_3}b, reading out either system Q1Q2 (red) or system Q3Q4 (green). This measurement shows that the states of Q2 and Q3 are swapped, but imperfections in the readout and initialisation and in the calibration of the resonant exchange pulse result in a small residual amplitude on Q3. 
    
    
     \begin{figure*}[!t]
	\includegraphics[width = \textwidth]{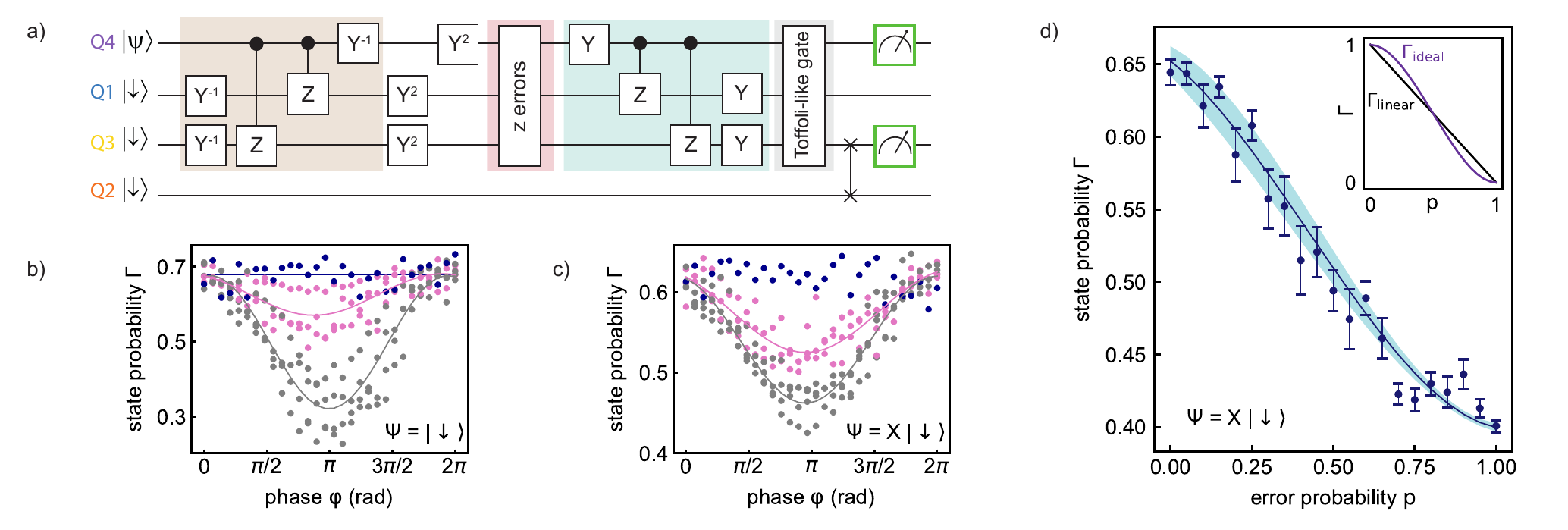}%
	\caption{\textbf{Three-qubit phase flip code.} (a) Implementation of the three-qubit phase flip code. Q4 serves as data qubit, Q1 and Q3 as ancilla qubits, and Q2 as reset qubit for Q3. The correction step is implemented with the Toffoli-like gate as shown in Figure \ref{fig:figure_3}e. The data qubit is read out with readout system Q3Q4, which makes it necessary to reset Q3 using a resonant SWAP operation with Q2. (b,c) Results of the phase error correction while introducing a phase error Z($\varphi$) on none of the qubits (dark blue), on one qubit (pink) or on two or all three qubits (gray). The initial state of the data qubit is $\ket{\downarrow}$ in (b) and X$\ket{\downarrow}$ in (c). (d) Phase-error correction by introducing phase errors Z($\varphi$ = $\pi$) with probability $p$. The data qubit is prepared to the state X$\ket{\downarrow}$ and through single qubit gates projected to the $\ket{\downarrow}$ state for readout. Plotted is the state probability ($\Gamma$). The results are fitted using a model which takes into account the readout and reset errors (see Supplementary Information section \RomanNumeralCaps{8}). The standard deviation of the fit is indicated by the light blue area. The inset shows the s-curve for ideal phase flip correction ($\Gamma_\text{ideal}$) and the linear line for no phase flip correction ($\Gamma_\text{linear}$).}
	\label{fig:figure_4}
    \end{figure*}
    
    Figure \ref{fig:figure_3}c shows the circuit diagram to demonstrate the controlled-$\text{S}^{-1}$ ($\text{CS}^{-1}$) gate. The calibration of the $\text{CS}^{-1}$ gate is similar to the CZ gate (see Supplementary Information section \RomanNumeralCaps{2}), however, for the $\text{CS}^{-1}$ gate the exchange pulse is calibrated to give a phase difference of -$\pi/2$ between the experiments with and without a preparation pulse on the control qubit. This is demonstrated in Figure \ref{fig:figure_3}d, where the results of a Ramsey type experiment are shown with (blue) and without (black) an $\text{X}^2$ pulse on the control qubit Q1. 
    
    The Toffoli gate is a three-qubit gate, also called Controlled-Controlled-NOT gate. In the three-qubit phase flip code, the combination of the decoding step and the Toffoli gate performs the majority vote. A resonant i-Toffoli was demonstrated in previous work \cite{hendrickx_four-qubit_2021}, which would be the fastest implementation when working in a regime where the exchange is on \cite{gullans_protocol_2019}. However, the qubit states are also strongly sensitive to noise in this regime. Here, we implement a Toffoli-like gate composed of CZ and C$\text{S}^{-1}$ gates (Figure \ref{fig:figure_3}e). This is equal to a Toffoli gate up to single and two-qubit rotations on the control qubits \cite{barenco_elementary_1995,smolin_five_1995}, which are irrelevant in the protocol under study \cite{taminiau_universal_2014}. The matrix representing this gate is shown in Supplementary Information section \RomanNumeralCaps{5}. We test the Toffoli-like gate by applying it to different input states, as shown in Figure \ref{fig:figure_3}e. Here a Rabi pulse X($\theta$) is applied to control qubit Q1 and the state of the target qubit Q4 is measured using readout system Q3Q4. Figure \ref{fig:figure_3}f shows the result with (pink) and without (black) an additional preparation pulse on the second control qubit, Q3. If neither of the control qubits is in the $\ket{\uparrow}$ state (when X($\theta$ = 0)) or when only one control qubit is in the $\ket{\uparrow}$ state (when X($\theta$ = $\pi$) on Q1), the target qubit remains in the $\ket{\downarrow}$ state. Only when both control qubits are in the $\ket{\uparrow}$ state, the target qubit flips. The result of a similar experiment, where a Rabi pulse X($\theta)$ is applied to the other control qubit, Q3, is shown in the Supplementary Information section \RomanNumeralCaps{5}. By applying a Rabi pulse X($\theta$) on Q1 (Q3), it is shown that this implementation of the Toffoli-like gate works for all X($\theta)\ket{\downarrow}$ of Q1 (Q3). 
    
    We now turn to the three-qubit phase flip correction code, of which the circuit diagram is shown in Figure \ref{fig:figure_4}a. The four qubit system is initialized to the $\ket{\downarrow\downarrow\downarrow\downarrow}$ state, after which we prepare the data qubit, Q4, in a state $\ket{\Psi}$. After the encoding step (beige), refocusing pulses are applied on all three qubits. The errors are implemented by either sweeping the phase Z($\varphi$) or by applying a full phase flip Z($\varphi$ = $\pi$) with probability $p$ (soft red). Subsequently, the qubits are decoded (turquoise). The correction step (grey) is implemented with the Toffoli-like gate shown in Figure \ref{fig:figure_3}e. The data qubit state is projected through single qubit gates to $\ket{\downarrow}$, the states of Q3 and Q2 are swapped and finally the data qubit, Q4, is read out using readout system Q3Q4. 
    
    This quantum error correction code corrects for a full phase flip as well as an arbitrary Z($\varphi$) rotation on a single qubit. When a phase error Z($\varphi$) occurs on a qubit, it is in a superposition of being in the correct state (with the ancillas indicating as such) and a state with a phase error (with one or both ancillas being flipped), and the Toffoli-like gate will be able to correct this superposition state. Figure \ref{fig:figure_4}b shows the state probability ($\Gamma$) (i.e. the chance that the data qubit is successfully rotated back to the $\ket{\downarrow}$ state) for errors implemented by sweeping the phase Z($\varphi$). This error is applied to one qubit at a time (for Q4, Q1 and Q3 individually, pink), to combinations of two qubits (Q4\&Q1, Q4\&Q3 and Q1\&Q3, grey) and to all three qubits at once (grey). For comparison, the result is also shown when the phase flip code is performed without sweeping the phase on any of the qubits (dark blue). These experiments are performed for two different input states of the data qubit, a basis state ($\ket{\Psi}$ = $\ket{\downarrow}$) and a superposition state ($\ket{\Psi}$ = X$\ket{\downarrow}$), shown in Figure \ref{fig:figure_3}b and \ref{fig:figure_3}c respectively. Only when the data qubit is prepared in a superposition state, does the encoding step entangle the data qubit with the two ancilla qubits. One expects that when sweeping the phase Z($\varphi$) on one of the qubits, the error is corrected and the result is a constant high $\Gamma$. For sweeping the phase on two or three qubits simultaneously, it is expected that the error is not corrected and $\Gamma$ varies from high to low and back. It is apparent from the results in Figure \ref{fig:figure_4}b and \ref{fig:figure_4}c that for both input states the single-qubit errors are not corrected perfectly. This is due to unintentional errors, i.e. errors occurring in the encoding, decoding and correction steps of the algorithm. These errors are caused by decoherence of the qubits, residual exchange  between the qubits (see Supplementary Information section \RomanNumeralCaps{6}), cross talk \cite{lawrie_simultaneous_2021, heinz_crosstalk_2021} and imperfect two-qubit gates. When comparing the results for the different input states of the data qubit, it becomes clear that for an input state $\ket{\downarrow}$ of the data qubit the visibility is higher and the correction of the single-qubit errors is more successful. We ascribe this improvement to the decreased time during which any unintentional errors can affect the data qubit when it starts in a basis state, and the fact that is less sensitive to imperfections in the two-qubit gates. 
    
    We also study the three-qubit error phase flip code by inserting a full phase flip Z($\varphi$ = $\pi$) with probability $p$. The data qubit starts in the state $\ket{\Psi}$ = $X\ket{\downarrow}$ and before the data qubit is measured, we project the state to $\ket{\downarrow}$ by applying corresponding single qubit gates. In Figure \ref{fig:figure_4}d the state probability $\Gamma$ is plotted against the probability of implementing a phase flip ($p$). In contrast to some experiments \cite{reed_realization_2012, taminiau_universal_2014}, where error rates are modeled by averaging over individual measurements of all possible phase errors, here we statistically perform large sequences where errors on all three qubits may occur at random instances. This procedure requires significantly more data and results in larger error bars, but it does capture the realistic scenario in which phase errors may occur.
    Ideally, the data is described by $\Gamma_{\rm{ideal}}(p) = 1 - 3p^2 +2p^3$ \cite{nielsen_quantum_2000}, which is the probability for having at most a single error. The ideal function shows a modest improvement of the state probability for $p < 0.5$, compared to the linear line $\Gamma_{\rm{linear}}(p) = 1-p$ expected for no error correction (see inset Figure \ref{fig:figure_4}d). While the data in Figure \ref{fig:figure_4}d follows the overall trend, there are also some interesting differences. We first note that since the data qubit starts in the $\ket{\Psi}$ = $X\ket{\downarrow}$ state, the data qubit is very sensitive to unintentional errors, resulting in a reduced visibility. Second, the data in Figure \ref{fig:figure_4}d does not show a trend symmetric around $\Gamma(p=0.5)$. This asymmetry can be caused by errors in the reset of Q3 combined with asymmetry in the readout scheme (illustrated in the Supplementary Information section \RomanNumeralCaps{1}). For example, when the reset of Q3 using the SWAP is imperfect, the error that is introduced is not random, but depends on the history of Q3. Due to the asymmetry of the readout, an imperfect reset of Q3 affects the measurement results of zero or a single intentional error differently than the measurement results for two or three intentional errors. These considerations can be taken into account using the fit function: $\Gamma(p) = b + a (0.95 - 1.73\epsilon\, p - 2.79\, p^2 + 3.9\epsilon\, p^2 + 1.86 p^3 - 2.17 \epsilon\, p^3)$, where $a$ is the visibility, $b$ the offset and $\epsilon$ the error parameter modeling asymmetry (See Supplementary Information section \RomanNumeralCaps{7} for the derivation). We obtain the fit parameters $a = 0.272 \pm 0.007$, $b = 0.394 \pm 0.003$ and $\epsilon = 0.37 \pm 0.13$. As expected $a$ and $b$ reflect that the visibility is reduced and that the offset is significant. We note that if $\epsilon=0$, the expected shape of $\Gamma_{\text{ideal}}$ is recovered, meaning that $\Gamma$ is improved for $p<0.5$ when comparing to $\Gamma_{\text{linear}}$ with similar visibility. For $\epsilon = 0.37$, as fitted here, the state probability still shows a small improvement for $p<0.27$, compared to $\Gamma_{\text{linear}}$.
    
    After two decades of quantum computation with quantum dots \cite{loss_quantum_1998}, it is now possible to implement rudimentary error correction circuits. We have implemented a two-qubit phase flip code and confirmed that by applying an echo pulse to the ancilla qubit we can preserve the state of the data qubit. We have demonstrated a resonant SWAP gate in germanium and have implemented a Toffoli-like gate using CZ and $\text{CS}^{-1}$ gates. Utilizing these gates has allowed us to implement a three-qubit phase flip code. Though scaling quantum dots in two dimensions and readout using Pauli spin blockade are central aspects in virtually all semiconductor qubit architectures \cite{vandersypen_interfacing_2017}, we have also observed that they affect the quantum gate compilation as well as the correction itself. While formidable improvements will have to be made to obtain fault-tolerant operation, we envision that the capability to test tailored quantum algorithms in real devices will serve as a crucial link in developing scalable quantum technology.
    
\section*{Supplementary Information}	
The Supplementary Information contains details on the calibration of the CZ and C$\text{S}^{-1}$ gates (section \RomanNumeralCaps{1}), more information about the latched PSB readout protocol (section \RomanNumeralCaps{2}), additional data on the two-qubit phase flip code (section \RomanNumeralCaps{3}), details on the calibration of the resonant SWAP gate (section \RomanNumeralCaps{4}), additional data on the Toffoli-like gate (section \RomanNumeralCaps{5}), data on the residual exchange (section \RomanNumeralCaps{6}), and the derivation of the fit function for the results of the three-qubit phase flip (section \RomanNumeralCaps{7}). 

\section*{Acknowledgements}
We thank all the members of the Veldhorst group for inspiring discussions. M. V. acknowledges support through two projectruimtes and a Vidi grant, associated with the Netherlands Organization of Scientific Research (NWO), and an ERC Starting Grant. Research was sponsored by the Army Research Office (ARO) and was accomplished under Grant No. W911NF- 17-1-0274. M. Rispler and B. M. T. are supported by QuTech NWO funding 2020-2024 – Part I “Fundamental Research” with project number 601.QT.001-1. M. Rispler acknowledges support by the EU Quantum Technology Flagship grant AQTION under Grant Agreement number 820495. The views and conclusions contained in this document are those of the authors and should not be interpreted as representing the official policies, either expressed or implied, of the Army Research Office (ARO), or the U.S. Government. The U.S. Government is authorized to reproduce and distribute reprints for Government purposes notwithstanding any copyright notation herein.

\section*{Data Availability}
The data that support the findings of this study will be openly available at 4TU.ResearchData.

\section*{Competing Interests}
The authors declare no competing interests. Correspondence should be addressed to M.V. (M.Veldhorst@tudelft.nl).

\end{document}